\documentstyle[epsfig]{article}
\begin{document}
\def\be{\begin{equation}}
\def\ee{\end{equation}}
\def\bea{\begin{eqnarray}}
\def\eea{\end{eqnarray}}
\def\OO#1{{\cal O}(c^{-#1})}

\begin{center}
{\large \bf Tests of relativity using a microwave resonator}\\
\vspace{5mm}
Peter Wolf$^{1,2}$, S\'ebastien Bize$^2$, Andr\'e Clairon$^2$, Andr\'e N. Luiten$^3$, Giorgio Santarelli$^2$, Michael E. Tobar$^3$\\
\vspace{5mm}
{\it $^1$Bureau International des Poids et Mesures, Pavillon de Breteuil, 92312 S\`evres Cedex, France \\
$^2$BNM-SYRTE, Observatoire de Paris, 61 Av. de l'Observatoire, 75014 Paris, France \\
$^3$University of Western Australia, School of Physics, Nedlands 6907 WA, Australia}
\end{center}

\begin{abstract}
The frequencies of a cryogenic sapphire oscillator and a hydrogen maser are compared to set new constraints on a possible violation of Lorentz invariance. We determine the variation of the oscillator frequency as a function of its orientation (Michelson-Morley test) and of its velocity (Kennedy-Thorndike test) with respect to a preferred frame candidate. We constrain the corresponding parameters of the Mansouri and Sexl test theory to $\delta - \beta + 1/2 = (1.5\pm 4.2) \times 10^{-9}$ and $\beta - \alpha - 1 = (-3.1\pm 6.9) \times 10^{-7}$ which is equivalent to the best previous result for the former and represents a 30 fold improvement for the latter.
\end{abstract}

PACS number(s): 03.30.+p, 06.30.Ft
\vspace{5mm}
\\
The Einstein equivalence principle (EEP) is at the heart of special and general relativity \cite{Will} and a cornerstone of modern physics. One of the constituent elements of EEP is Local Lorentz invariance (LLI) which, loosely stated, postulates that the outcome of any local test experiment is independent of the velocity of the (freely falling) apparatus. The central importance of this postulate in modern physics has motivated tremendous work to experimentally test LLI \cite{Will}. Additionally, nearly all unification theories (in particular string theory) violate the EEP at some level \cite{Damour1} which further motivates experimental searches for such violations of the universality of free fall \cite{Damour2} and of Lorentz invariance \cite{Kosto1, Kosto2}.

	The vast majority of modern experiments that test LLI rely essentially on the stability of atomic clocks and macroscopic resonators \cite{Brillet, KT, Hils, Schiller}, therefore improvements in oscillator technology have gone hand in hand with improved tests of LLI. Our experiment is no exception, the 30 fold improvement being a direct result of the excellent stability of our cryogenic sapphire oscillator. Additionally its operation at a microwave frequency allows a direct comparison to a hydrogen maser which provides a highly stable and reliable reference frequency.

    Numerous test theories that allow the modeling and interpretation of experiments that test LLI have been developed. Kinematical frameworks \cite{Robertson, MaS} postulate a simple parametrisation of the Lorentz transformations with experiments setting limits on the deviation of those parameters from their special relativistic values. A more fundamental approach is offered by theories that parametrise the coupling between gravitational and non-gravitational fields (TH$\epsilon\mu$ \cite{LightLee, Will, Blanchet} or $\chi$g \cite{Ni} formalisms) which allow the comparison of experiments that test different aspects of the EEP. Finally, formalisms based on string theory \cite{Damour1, Damour2, Kosto1} have the advantage of being well motivated by theories of physics that are at present the only candidates for a unification of gravity and the other fundamental forces of nature.

    Owing to their simplicity the kinematical frameworks of \cite{Robertson, MaS} have been widely used to model and interpret many previous experiments testing LLI \cite{Brillet, Hils, Schiller, Riis, WP}. In order to compare our results to those experiments we will follow this route in the present work (an analysis of our experiment in the light of other test theories being relegated to a future publication). Those frameworks postulate generalized transformations between a preferred frame candidate $\Sigma(T,{\bf X})$ and a moving frame ${\rm S}(t,{\bf x})$ where it is assumed that in both frames coordinates are realized by identical standards (e.g. hydrogen masers for the time coordinates). We start from the transformations of \cite{MaS} (in differential form) for the case where the velocity of S as measured in $\Sigma$ is along the positive X-axis, and assuming Einstein synchronization in S (we will be concerned with signal travel times around closed loops so the choice of synchronization convention can play no role):

\be
dT = {1\over a}\left(dt+{vdx\over c^2}\right); dX = {dx\over b}+{v\over a}\left(dt+{vdx\over c^2}\right); dY = {dy\over d}; dZ = {dz\over d}
\ee
with $c$ the velocity of light in vacuum in $\Sigma$. Using the usual expansion of the three parametrs $(a \approx 1+\alpha{v^2/c^2} + \OO4; b \approx 1+\beta{v^2/c^2} + \OO4; d \approx 1+\delta{v^2/c^2} + \OO4)$, setting $c^2dT^2=dX^2+dY^2+dZ^2$ in $\Sigma$, and transforming according to (1) we find the coordinate travel time of a light signal in S:

\be
dt={dl\over c}\left(1-\left(\beta -\alpha -1 \right){v^2\over c^2} - \left({1\over 2}-\beta +\delta \right){\rm sin}^2\theta{v^2\over c^2}\right)+\OO4
\ee
where $dl = \sqrt{dx^2+dy^2+dz^2}$ and $\theta$ is the angle between the direction of light propagation and the velocity {\bf v} of S in $\Sigma$. In special relativity $\alpha = -1/2; \beta = 1/2; \delta = 0$ and (1) reduces to the usual Lorentz transformations. Generally, the best candidate for $\Sigma$ is taken to be the frame of the cosmic microwave background (CMB) \cite{Fixsen, Lubin} with the velocity of the solar system in that frame taken as $v_\odot \approx 377$ km/s, decl. $\approx -6.4 ^\circ $, $RA \approx 11.2$h.

    Michelson-Morley type experiments \cite{MM, Brillet} determine the coefficient $P_{MM} = (1/2-\beta +\delta)$ of the direction dependent term. At present the most stringent limit on that parameter is $|P_{MM}| \leq 3.2 \times 10^{-9}$ \cite{Brillet, Brilletcom} determined over 23 years ago in an outstanding experiment whose uncertainty has not been approached since. Our experiment is the first one that can confirm that result with roughly equivalent uncertainty $(4.2 \times 10^{-9})$. Kennedy-Thorndike experiments \cite{KT, Hils, Schiller} measure the coefficient $P_{KT} = (\beta -\alpha -1)$ of the velocity dependent term. The most stringent limit on $|P_{KT}|$ has been recently improved from \cite{Hils} by a factor 3 to $|P_{KT}| \leq 2.1 \times 10^{-5}$  \cite{Schiller}. We improve this result by a factor of 30 to $|P_{KT}| \leq 6.9 \times 10^{-7}$. Finally clock comparison and Doppler experiments measure $\alpha$, currently limiting it to $|\alpha + 1/2| \leq 8 \times 10^{-7}$ \cite{Riis, WP, Isaakcom}. The three types of experiments taken together then completely characterize any deviation from Lorentz invariance in this particular test theory.

    Our cryogenic oscillator consists of a sapphire crystal of cylindrical shape operating in a whispering gallery mode (see fig. 1 for a schematic drawing and \cite{Chang, Mann} for a detailed description). Its coordinate frequency can be expressed by  $\nu = m/t_c$ where $t_c$ is the coordinate travel time of a light signal around the circumference of the cylinder (of radius $r$) and $m$ is a constant. From (2) the relative frequency difference between the sapphire oscillator and the hydrogen maser (which realizes coordinate time in S) is

\be
{\Delta \nu (t) \over \nu_0} = P_{KT}{v(t)^2\over c^2} + P_{MM}{v(t)^2\over c^2}{1 \over 2\pi}\int_0^{2 \pi}{\rm sin}^2\theta (t,\varphi ) d\varphi +\OO3
\ee
where $\nu_0 = m/(2\pi r/c)$, $v(t)$ is the (time dependent) speed of the lab in $\Sigma$, and $\varphi$ is the azimuthal angle of the light signal in the plane of the cylinder. The periodic time dependence of $v$ and $\theta$ due to the rotation and orbital motion of the Earth with respect to the CMB frame allow us to set limits on the two parameters in (3) by adjusting the periodic terms of appropriate frequency and phase (see \cite{Mike} for calculations of similar effects for several types of oscillator modes). Given the limited durations of our data sets ($\leq$ 9.5 days) the dominant periodic terms arise from the Earth's rotation, so retaining only those we have ${\bf v}(t) = {\bf u}+{\bf \Omega} \times {\bf R}$ with ${\bf u}$ the velocity of the solar system with respect to the CMB, ${\bf \Omega}$ the angular velocity of the Earth, and ${\bf R}$ the geocentric position of the lab. We then find after some calculation

\be
\begin{array}{cl}
\Delta \nu / \nu_0 &= P_{KT}(H{\rm sin}\lambda )\\
\ &
+ P_{MM}(A{\rm cos}\lambda + B{\rm cos}(2\lambda)+C{\rm sin}\lambda+D{\rm sin}\lambda{\rm cos}\lambda+E{\rm sin}\lambda{\rm cos}(2\lambda))
\end{array}
\ee
where $\lambda =\Omega t + \phi$, and A-E and $\phi$ are constants depending on the latitude and longitude of the lab $(\approx 48.7 ^\circ$N and $2.33 ^\circ$E for Paris). Numerically $H \approx -2.6 \times 10^{-9}$, $A \approx -8.8 \times 10^{-8}$, $B \approx 1.8 \times 10^{-7}$, C-E of order $10^{-9}$. We note that in (4) the dominant time variations of the two combinations of parameters are in quadrature and at twice the frequency which indicates that they should decorelate well in the data analysis allowing a simultaneous determination of the two (as confirmed by the correlation coefficients below). Adjusting this simplified model to our data we obtain results that differ by less than 10\% from the results presented below that were obtained using the complete model ((3) including the orbital motion of the Earth).

	The cryogenic sapphire oscillator (CSO) is compared to a commercial (Datum Inc.) active hydrogen maser whose frequency is also regularly compared to caesium and rubidium atomic fountain clocks in the laboratory \cite{Bize}. Both oscillators are operated in temperature controlled rooms, with the temperature sensitive electronics mounted on an actively temperature stabilized panel. The CSO resonant frequency at 11.932 GHz is compared to the 100 MHz output of the hydrogen maser. The maser signal is multiplied up to 12 GHz of which the CSO signal is subtracted. The  remaining $\approx$ 67 MHz signal is mixed to a synthesizer signal at the same frequency and the low frequency beat at $\approx$ 64 Hz is counted, giving access to the frequency difference between the maser and the CSO. The instability of the comparison chain has been measured and does not exceed a few parts in $10^{16}$. The typical stability of the measured CSO - maser frequency after removal of a linear frequency drift is shown in fig. 1.

    Our experimental data consists of seven sets of measured values of $\Delta \nu / \nu_0$ of varying length (2.8 to 9.5 days, 37 days in total) taken in Nov./Dec. 2001 and Mar./Apr./Sep. 2002 (see fig. 2). The sampling times $\tau _0$ are generally 100 s except for the first two sets for which $\tau_0 = 12$ s and $5000$ s respectively. To analyze our data we simultaneously adjust an offset and a rate (natural frequency drift, typically $\approx 2 \times 10^{-18}$ s$^{-1}$) per data set and the two parameters of the model (3), a total of 16 parameters. In the model (3) we take into account the rotation of the Earth and the Earth's orbital motion, the latter contributing little as any constant or linear terms over the durations of the individual data sets are absorbed by the adjusted offsets and rates. To ensure homogeneity between the data sets we average all sets to $\tau_0 = 5000$ s.

    We first carry out an ordinary least squares (OLS) adjustment obtaining $|P_{MM}| = (-2.8\pm 3.6) \times 10^{-9}$ and $|P_{KT}| = (-4.3\pm 2.8) \times 10^{-7}$. The correlation coefficient between the two parameters is less than 0.01 indicating that the two are indeed well decorelated and can be determined simultaneously. All other correlation coefficients between either of the two parameters and the 14 adjusted offsets and rates are less than 0.07. We note, however, that the residuals have a significantly non-white behavior as one would expect from the slope of the Allan deviation of fig. 1. The power spectral density (PSD) of the residuals when fitted with a power law model of the form $S_y(f)=kf^\mu$ yields typically $\mu \approx -1.5$. In the presence of non-white noise OLS is not the optimal regression method \cite{lss, Draper} as it can lead to significant underestimation of the parameter uncertainties \cite{lss}.

    An alternative method is weighted least squares (WLS) \cite{Draper} which allows one to account for non-random noise processes in the original data by pre-multiplying both sides of the design equation (our equation (3) plus the 14 offsets and rates) by a weighting matrix containing off diagonal elements. To determine these off diagonal terms we first carry out OLS and adjust the $S_y(f)=kf^\mu$ power law model to the PSD of the post-fit residuals determining a value of $\mu$ for each data set. We then use these $\mu$ values to construct a weighting matrix following the method of fractional differencing described, for example, in \cite{lss}. The WLS regression yields $|P_{MM}| = (1.5\pm 3.1) \times 10^{-9}$ and $|P_{KT}| = (-3.1\pm 3.7) \times 10^{-7}$ ($1\sigma$ uncertainties), with the correlation coefficient between the two parameters less than 0.01 and all other correlation coefficients $< 0.07$. The best fit power law of the WLS residuals is now compatible with $\mu \approx 0$. The inset of fig. 2 shows the last data set (Sep. 2002) and the best fit model after pre-multiplication of both by the weighting matrix.

    The two methods give similar results for the two parameters but we consider WLS as more reliable and take its outcome as the final results of our statistical analysis.

	Systematic effects at diurnal or semi-diurnal frequencies with the appropriate phase could partially cancel a putative sidereal signal as our total data span ($\approx$ 293 days) allows only partial separation of the diurnal from the sidereal signal. The statistical uncertainties of $P_{MM}$ and $P_{KT}$ obtained from the WLS fit above correspond to sidereal and semi-sidereal terms (from (4)) of $\approx 1 \times 10^{-15}$ and $\approx 6 \times 10^{-16}$ respectively so any systematic effects exceeding these limits need to be taken into account in the final uncertainty. We expect the main contributions to such effects to arise from temperature, pressure and magnetic field variations that would affect the hydrogen maser, the CSO and the associated electronics, and from tilt variations of the CSO which are known to affect its frequency. Measurements of the tilt variations of the CSO show amplitudes of 4.6 $\mu$rad and 1.6 $\mu$rad at diurnal and semi-diurnal frequencies. To estimate the tilt sensitivity we have intentionally tilted the oscillator by $\approx$ 5 mrad off its average position which led to relative frequency variations of $\approx 3 \times 10^{-13}$ from which we deduce a tilt sensitivity of $\approx 6 \times 10^{-17} \mu$rad$^{-1}$. This value corresponds to a worst case scenario as we expect a quadratic rather than linear  frequency variation for small tilts around the vertical. Even with this pessimistic estimate diurnal and semi-diurnal frequency variations due to tilt do not exceed $3 \times 10^{-16}$ and $1 \times 10^{-16}$ respectively and are therefore negligible with respect to the statistical uncertainties. The temperature sensitive electronics were mounted on an actively temperature controlled panel reducing temperature fluctuations by about one order of magnitude. Temperature measurements of the CSO lab and the electronics panel taken during some of the experimental runs show room temperature variations with amplitudes of 0.3 $^\circ$C and 0.1 $^\circ$C for the diurnal and semi-diurnal components which are reduced to 0.04 $^\circ$C and 0.01 $^\circ$C on the panel. The hydrogen maser is kept in a dedicated clock room with temperature variations below the above values. Measurements of magnetic field, temperature and atmospheric pressure in that room and the maser sensitivities as specified by the manufacturer allow us to exclude any systematic effects on the maser frequency that would exceed the statistical uncertainties above. Switching off the temperature stabilization of the electronics panel shows no discernible effect so we are confident in excluding any systematic effects from that source. When heating and cooling the CSO lab by $\approx 3^\circ$C we see frequency variations of $\approx 5 \times 10^{-15}$ per $^\circ$C. From the temperature measurements during the experimental runs we therefore deduce a total diurnal and semi-diurnal effect of $\approx 1.5 \times 10^{-15}$ and $\approx 5.0 \times 10^{-16}$ respectively. We assume the pessimistic scenario where all of the diurnal and semi-diurnal effect is present at the neighboring sidereal and semi-sidereal frequencies which leads (from (3)) to uncertainties from systematic effects of $\pm 5.8 \times 10^{-7}$ on $P_{KT}$ and $\pm 2.8 \times 10^{-9}$ on $P_{MM}$. We note that the phase of the perturbing systematic signal will vary over the course of our measurements due to natural causes (meteorology, daytime changes etc.) and to the sidereal/diurnal frequency difference so our final uncertainties given below are the quadratic sums of the above values and the statistical uncertainties from the WLS adjustment.

	In summary, we have reported an experimental test of Lorentz invariance that simultaneously constrains two combinations of the three parameters of the Mansouri and Sexl test theory (previously measured individually by Michelson-Morley and Kennedy-Thorndike experiments). Our experiment limits $|\delta - \beta + 1/2|  \leq 4.2\times 10^{-9}$ which for the first time confirms the 1979 value of \cite{Brillet}, and $|\beta - \alpha - 1| \leq 6.9\times 10^{-7}$ which improves the best present limit \cite{Schiller} by a factor of 30. As a result, the Lorentz transformations are confirmed in this particular test theory with an overall uncertainty of $\leq 8 \times 10^{-7}$ limited now by the determination of $\alpha$ from Doppler and clock comparison experiments \cite{Riis, WP, Isaakcom}. This is likely to be improved in the coming years by experiments such as ACES (Atomic Clock Ensemble in Space \cite{ACES}) that will compare ground clocks to clocks on the international space station aiming at a 10 fold improvement on the determination of $\alpha$. We hope to improve our experiment by a better characterization and control of systematic effects (in particular due to temperature) and by further data integration. This should allow us to improve our limits by another factor 2 or 3 in the near future. 

\vspace{5mm}

\noindent{\bf Acknowledgements:}
\noindent Helpful discussions with Christophe Salomon and G\'erard Petit are gratefully acknowledged as well as financial support by the Australian Research Council and CNES. P.W. is supported by CNES research grant 793/02/CNES/XXXX.
\vspace{5mm}

%%\noindent{\bf Figure captions}
\begin{center}
\includegraphics[width=140mm,angle=0]{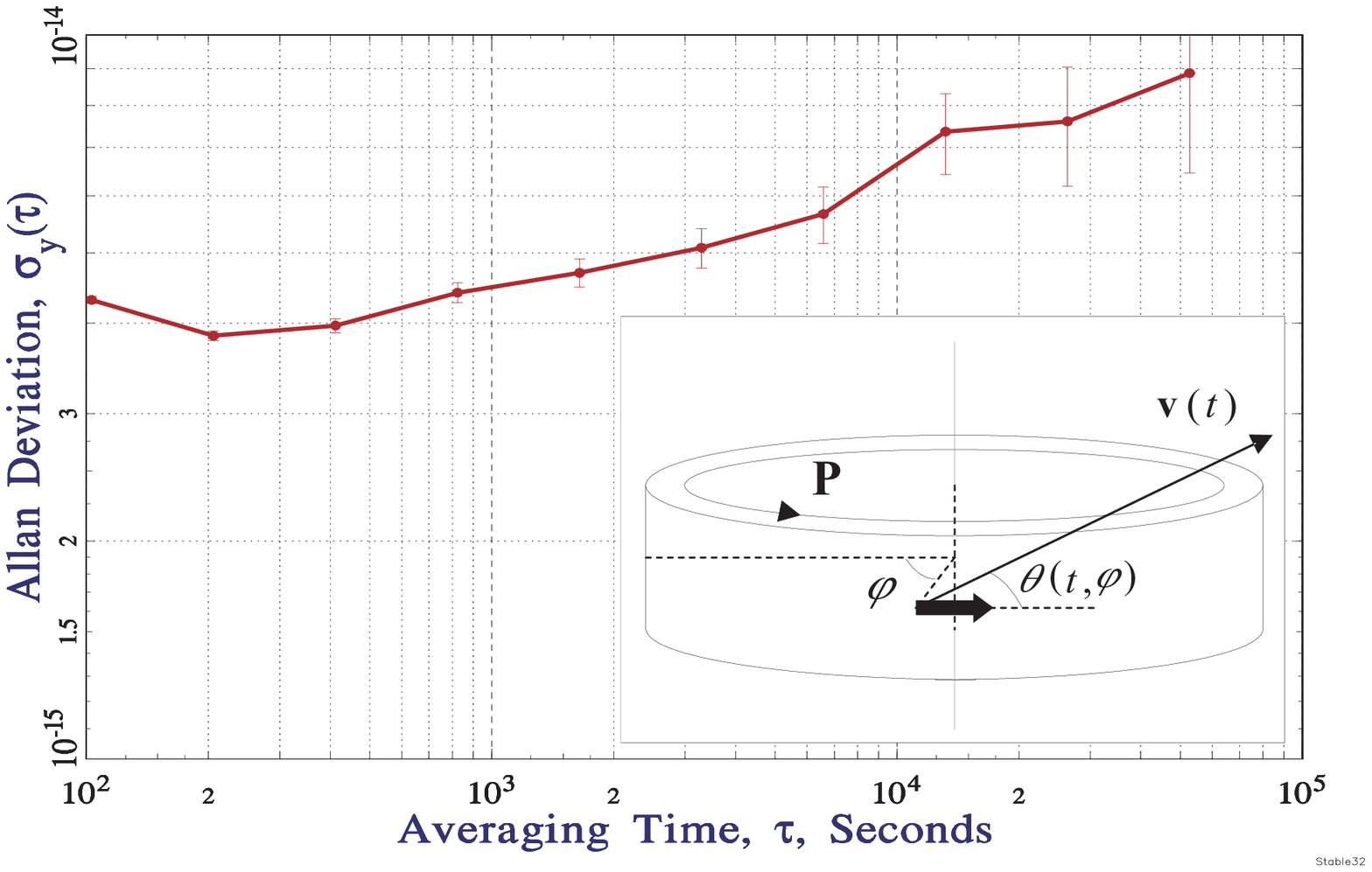}
\end{center}

\noindent Fig. 1. Typical relative frequency stability of the CSO - H-maser difference after removal of a linear frequency drift. The inset is a schematic drawing of the cylindrical sapphire oscillator with the Poynting vector $\bf P$ in the whispering gallery (WG) mode, the velocity ${\bf v}(t)$ of the cylinder with respect to the CMB, and the relevant angles for a photon in the WG mode.

\begin{center}
\includegraphics[width=140mm,angle=0]{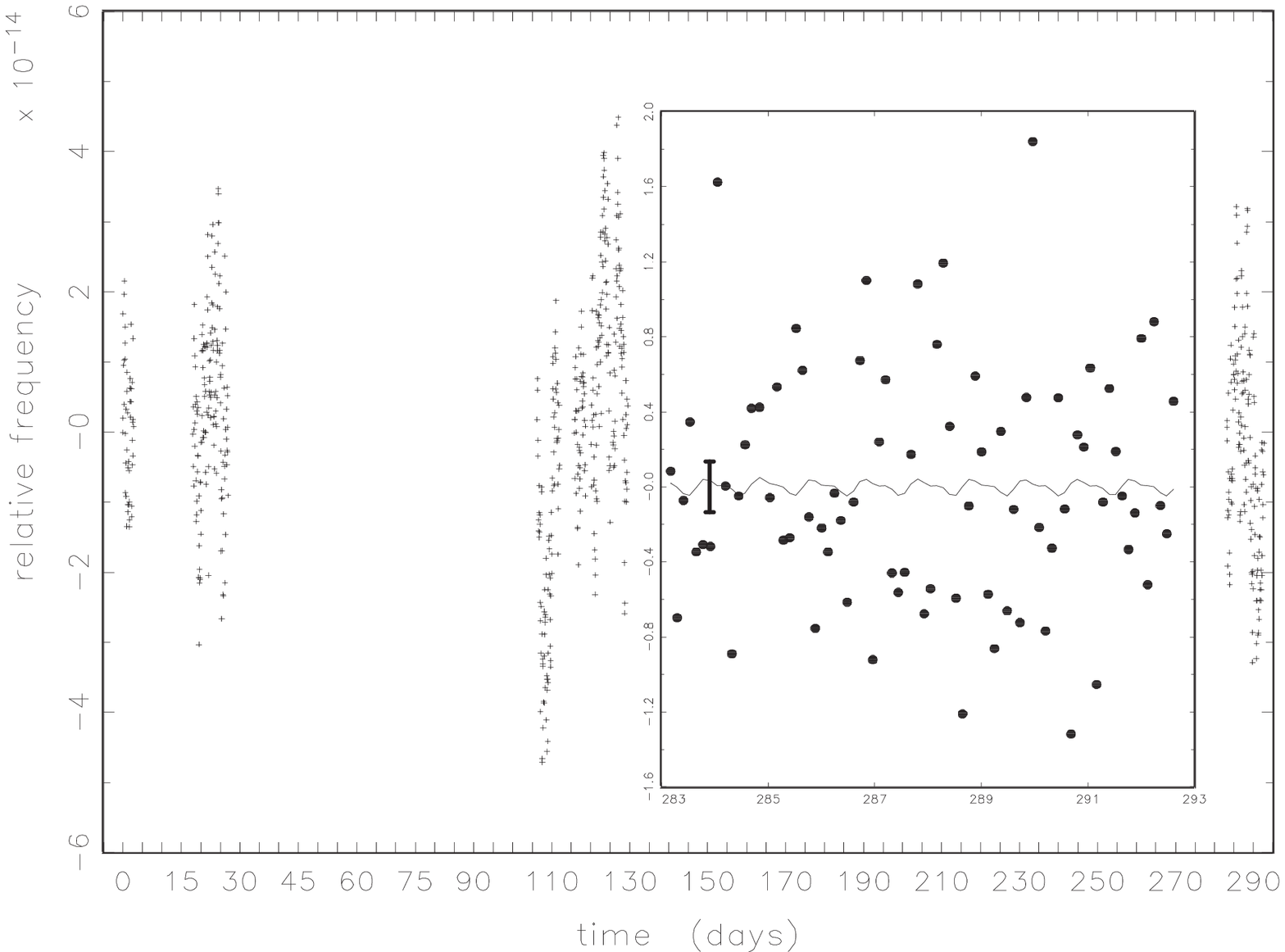}
\end{center}

\noindent Fig.2. Measured relative frequency difference between the CSO and the hydrogen maser after removal of the adjusted offsets and rates, and best fit model (3) of Lorentz invariance violation (solid line). The inset shows the last data set after pre-multiplication of the data and the model by the weighting matrix of the WLS adjustment (for presentation purposes data points were averaged by a factor 2). The error bar is the maximum signal amplitude within the final 1$\sigma$ uncertainties of the parameters.


\begin{thebibliography}{99}
\bibitem{Will} Will C.M., {\it Theory and Experiment in Gravitational Physics, revised edition},
Cambridge U. Press, (1993).
\bibitem{Damour1} Damour T., gr-qc/9711060 (1997).
\bibitem{Damour2} Damour T. and Polyakov A.M., Nucl.Phys. {\bf B423}, 532, (1994).
\bibitem{Kosto1} Colloday D. and Kostelecky V.A., Phys. Rev. {\bf D55}, 6760, (1997).
\bibitem{Kosto2} Bluhm R. et al., Phys. Rev. Lett. {\bf 88}, 9, 090801, (2002)
\bibitem{Brillet} Brillet A. and Hall J.L., Phys. Rev. Lett. {\bf 42}, 9, 549, (1979).
\bibitem{KT} Kennedy R.J. and Thorndike E.M., Phys. Rev. {\bf B42}, 400, (1932).
\bibitem{Hils} Hils D. and Hall J.L., Phys. Rev. Lett., {\bf 64}, 15, 1697, (1990).
\bibitem{Schiller} Braxmaier C. et al., Phys. Rev. Lett. {\bf 88}, 1, 010401, (2002).
\bibitem{Robertson} Robertson H.P., Rev. Mod. Phys. {\bf 21}, 378 (1949).
\bibitem{MaS} Mansouri R. and Sexl R.U., Gen. Rel. Grav. {\bf 8}, 497, 515, 809, (1977).
\bibitem{LightLee} Lightman A.P. and Lee D.L., Phys. Rev. {\bf D8}, 2, 364, (1973).
\bibitem{Blanchet} Blanchet L., Phys. Rev. Lett. {\bf 69}, 4, 559, (1992).
\bibitem{Ni} Ni W.-T., Phys. Rev. Lett. {\bf 38}, 301, (1977).
\bibitem{Fixsen} Fixsen D.J. et al., Phys. Rev. Lett. {\bf 50}, 620, (1983).
\bibitem{Lubin} Lubin et al., Phys. Rev. Lett. {\bf 50}, 616, (1983).
\bibitem{MM} Michelson A.A. and Morley E.W., Am. J. Sci., {\bf 34}, 333, (1887).
\bibitem{Brilletcom} The result of \cite{Brillet} is often quoted as $|P_{MM}| \leq 5 \times 10^{-9}$ (e.g. \cite{Hils, Schiller}). In \cite{Brillet} the result is given in terms of the Robertson parameters as $|g_2/g_1 - 1| \leq 5 \times 10^{-15}$ which is related to $P_{MM}$ by $g_2/g_1 - 1 = -P_{MM}(v/c)^2$ so we have $|P_{MM}| \leq 3.2 \times 10^{-9}$ for $v \approx$ 377 km/s.
\bibitem{Riis} Riis E. et al., Phys. Rev. Lett. {\bf 60}, 81, (1988).
\bibitem{WP} Wolf P. and Petit G., Phys. Rev. {\bf A56}, 6, 4405, (1997).
\bibitem{Isaakcom} We do not take into account the experiment briefly mentioned by G.R. Isaak in two paragraphs of his general review article (Phys. Bull. {\bf 21}, 255, 1970) often cited by other authors (e.g. \cite{Will, MaS, Hils, Schiller}) as, to our knowledge, the referenced publication (Isaak G.R. et al., 1970, Nature, to be published) has not been published in Nature or elsewhere. 
\bibitem{Chang} Chang S., Mann A.G. and Luiten A.N., Electron. Lett. {\bf 36}, 5, 480, (2000).
\bibitem{Mann} Mann A.G., Chang S. and Luiten A.N., IEEE Trans. Instrum. Meas. {\bf 50}, 2, 519, (2001).
\bibitem{Mike} Tobar M.E. et al., Phys. Lett. {\bf A300}, 33, (2002).
\bibitem{Bize} Bize S. et al., Proc. 6th Symp. on Freq. Standards and Metrology, World Scientific, (2002).
\bibitem{lss} Schmidt L.S., Metrologia {\bf 40}, in press, (2003).
\bibitem{Draper} Draper N.R. and Smith H., {\it Applied Regression Analysis}, Wiley, (1966).
\bibitem{ACES} Salomon C., et al., C.R. Acad. Sci. Paris, {\bf 2}, 4, 1313, (2001).
\end{thebibliography}
\end{document}